\documentclass[aps,prb,twocolumn,showpacs,superscriptaddress]{revtex4}
\usepackage{amssymb,amsmath}
\usepackage{graphicx}
\usepackage{multirow}
\usepackage{hyperref}
\usepackage{appendix}
\usepackage{mathtools}
\usepackage{xcolor}
\usepackage{siunitx}
\graphicspath{{./IMG/}}

\begin{document}
	
	\title{Combined magnetic and structural transition state in first-principles calculations.}
	
	\author{Samare Rostami*}
	
	\author{Xavier Gonze}
	\affiliation{European Theoretical Spectroscopy Facility, Institute of Condensed Matter and Nanosciences, Universit\'{e} catholique de Louvain, Chemin des \'{e}toiles 8, bte L07.03.01, B-1348 Louvain-la-Neuve, Belgium}
	
	\date{\today}
	
	\email{samare.rostami@gmail.com}
	\begin{abstract}
We present a generalized nudged elastic band (NEB) implementation within potential-based constrained density-functional theory (cDFT) 
for calculating transition pathways that simultaneously optimize atomic, lattice, and magnetic degrees of freedom.
Unlike standard fixed-cell NEB calculations restricted to atomic displacements, our framework consistently couples atomic forces, 
stresses, and spin torques. The method supports fixed- and variable-cell approaches for collinear and  
noncollinear magnetic configurations, by incorporating magnetic torques and generalized cell coordinates  within
the NEB optimization.  
To validate the methodology, we present benchmark calculations for representative magnetostructural systems.
Specifically, we investigate the bcc-to-hcp transformation in Fe and the antiferromagnetic-to-ferromagnetic
transition in FeRh using the variable-cell formalism. Furthermore, we examine spin-reversal processes in
two-dimensional CrI$_3$ to analyze the influence of spin-orbit coupling on transition paths and their associated energy barriers.
	\end{abstract}

	\maketitle
	
	\section{Introduction}
	%\section{Introduction}
	\label{Sec:intro}

\label{Sec:intro}

Many fascinating phenomena in condensed-matter physics are governed by activated transformations between 
competing metastable states, where both the atomic structure and magnetic order evolve cooperatively. 
Representative examples include magnetostructural phase transitions, pressure-induced magnetic 
transformations, magnetovolume effects, and magnetic switching 
\cite{Liu2025,Yang2026,He2024,Li2019,Han2023}. In these systems, the strong coupling between magnetic and 
structural degrees of freedom gives rise to significant lattice distortions and anisotropic cell variations 
during magnetic transitions \cite{Lewis2015}. These processes occur on complex multidimensional
potential-energy landscapes, making the identification of the minimum-energy path (MEP) connecting
the initial and final states essential for understanding transition mechanisms, activation barriers, 
and switching kinetics.

The nudged elastic band (NEB) method and its climbing-image extension are standard approaches for 
locating transition pathways in atomistic simulations \cite{Henkelman2000a,Henkelman2000b}. 
For solid-state transformations involving lattice deformations, the solid-state NEB (SS-NEB) 
method extends the configuration space to include both atomic positions and lattice vectors within 
a unified Euclidean framework \cite{Sheppard2012,Zarkevich2015,Gao2016,Fantasia2024}. 
Magnetic transitions introduce an additional geometric challenge because orientations of fixed-magnitude local magnetic 
moments evolve on the curved Riemannian manifold $\prod_i S^2$ rather than in Euclidean space. 
To address this, Bessarab \textit{et al.} developed the geodesic NEB (GNEB) method 
\cite{Bessarab2015,Spirit2019}, later complemented by geodesic minimum-mode following (GMMF) for direct 
saddle-point searches \cite{Schrautzer2025}. However, these work rely on spin Hamiltonians and neglect the explicit
self-consistent evolution of the electronic structure, limiting their applicability to itinerant magnets and
systems with significant charge redistribution.

Constrained-spin density functional theory (DFT) provides a first-principles route to magnetic transition 
states \cite{Dederichs1984,Kurz2004,Wu2005,Dudarev2015}. Recent developments, including the self-adaptive 
DeltaSpin framework \cite{Cai2023} and the potential-based constrained DFT (cDFT) formulation of Gonze 
\textit{et al.} \cite{Gonze2022}, have substantially improved the robustness of constrained magnetic 
calculations.
In the latter approach, magnetic constraints are incorporated directly into the 
potential-based self-consistency problem. The target magnetic moments are thereby treated as external 
parameters on the same footing as the atomic positions and lattice parameters, allowing spin torques, 
atomic forces, and stresses to be evaluated consistently. This formulation is therefore particularly 
well suited for coupled magnetostructural transition-state calculations.

The importance of coupling magnetic and structural relaxations has recently been emphasized. 
Lawrence \textit{et al.} showed that structural relaxation near magnetic transition states can reduce 
activation barriers by up to 20\% in antiferromagnetic systems \cite{Lawrence2023}. Tang \textit{et al.} 
subsequently introduced a magnetic string method based on DeltaSpin-constrained DFT calculations 
\cite{Tang2025}. Nevertheless, their approach neglects lattice degrees of freedom, restricting the 
calculations to fixed simulation cells and preventing a fully coupled magnetostructural description.

To our knowledge, no first-principles framework simultaneously optimizes atomic positions, lattice 
vectors, and magnetic moments along a transition pathway while maintaining full electronic self-consistency.
Existing approaches either decouple structural and magnetic optimizations or neglect macroscopic strain, 
which can lead to inaccurate transition mechanisms and energy barriers when lattice-spin coupling is strong.

In this work, we address this challenge by combining potential-based cDFT with a generalized 
transition-path methodology implemented in the ABINIT package. The resulting cDFT-NEB framework
enables the simultaneous optimization of atomic coordinates, lattice vectors, and non-collinear 
magnetic moments under full electronic self-consistency. 

We demonstrate the capabilities of the method using three representative applications: the AFM-to-FM
transition in FeRh, the coupled structural and magnetic transformation from bcc-FM to hcp-NM iron, and
noncollinear spin reversal in monolayer CrI$_3$. Together, these applications demonstrate that the framework consistently captures the coupled evolution of atomic, lattice, and magnetic degrees of freedom and allows the effects of cell relaxation and pressure on transition pathways and activation barriers to be assessed. Our work therefore provides a unified first-principles methodology for investigating transition mechanisms and activation barriers in strongly coupled magnetostructural materials.

The remainder of this paper is organized as follows. Section~II briefly reviews the potential-based 
constrained DFT formalism underlying the present implementation. Section~III presents the generalized 
cDFT-NEB methodology, including the treatment of atomic, lattice, and magnetic degrees of freedom as well as the optimization algorithms. Section~IV reports the computational details and benchmark applications
to FeRh, Fe, and monolayer CrI$_3$, demonstrating the capabilities of the proposed approach. 
Finally, Section~V summarizes the main conclusions and perspectives of this work.
%%%%%%%%%%%%%%%%%%%%%%%%%%%%%%%%%%%%%%%%%%%%%%%%%%%%%

\section{Method}
\label{Sec:method}

\subsection{Potential-based constrained density functional theory}

The present implementation is based on the potential-based constrained density-functional 
theory (cDFT) formalism introduced in Ref.~\onlinecite{Gonze2022}. In this framework, the screened 
Kohn-Sham potential is treated as the fundamental variational quantity, allowing charge and 
magnetic constraints to be incorporated directly within the self-consistent procedure. In 
contrast to conventional density-based approaches, the potential-based formulation places 
constrained variables, atomic coordinates, and lattice degrees of freedom on the same footing
within the variational formalism.

For a trial screened potential $u$, the associated Kohn-Sham equations are solved according to

\begin{equation}
(\hat{T}+\hat{u}) |\phi_{i,u}\rangle=
\varepsilon_{i,u}|\phi_{i,u}\rangle,
\end{equation}

where $\hat{T}$ is the kinetic-energy operator and
$|\phi_{i,u}\rangle$ are the Kohn-Sham orbitals associated with the trial potential $u$,   
and $\varepsilon_{i,u}$ denote the corresponding Kohn-Sham eigenvalues. 
The corresponding electronic density is obtained as

\begin{equation}
\rho^{\rm v}[u]({\bf r})
=
\sum_i f_i
\phi_{i,u}^*({\bf r})
\phi_{i,u}({\bf r}),
\end{equation}

with $f_i$ denoting the occupation numbers.

Within the potential-based formalism, the self-consistent electronic energy is recast as 
a minimization problem in the space of screened potentials,

\begin{equation}
E^{\rm SC}_{v_{\rm ext}}
=
\min_u
E^{\rm v}_{v_{\rm ext}}[u],
\end{equation}

with

\begin{eqnarray}
E^{\rm v}_{v_{\rm ext}}[u]
&=&
T[\{\phi_{i,u}\}]
+
\int
\rho^{\rm v}[u]({\bf r})
v_{\rm ext}({\bf r}) d{\bf r}
\nonumber\\
&+&
E_{\rm Hxc}[\rho^{\rm v}[u]],
\end{eqnarray}

where $v_{\rm ext}$ is the external ionic potential and $E_{\rm Hxc}$ contains the Hartree 
and exchange-correlation contributions.

The gradient of the functional with respect to the screened potential is given by~\cite{Gonze2022}

\begin{eqnarray}
\frac{\delta E^{\rm v}_{v_{\rm ext}}[u]}
{\delta u({\bf r})}
&=&
\int
\chi_0({\bf r},{\bf r}')
R^{\rm v}[u]({\bf r}')
d{\bf r}',
\end{eqnarray}

where $\chi_0({\bf r},{\bf r}')$ is the independent-particle susceptibility and

\begin{equation}
R^{\rm v}[u]
=
K^{\rm v}[u]-u
\end{equation}

is the potential residual. Here, $K^{\rm v}[u]$ represents the output screened potential 
obtained after solving the Kohn-Sham equations for a trial input potential $u$.
Self-consistency is reached when the input and output screened potentials become identical, 
corresponding to a vanishing residual $R^{\rm v}[u]=0$.

Within cDFT, constraints are introduced through Lagrange multipliers. For multiple 
constraints labelled by $I$, the augmented functional is written as

\begin{equation}
E^{+{\rm v}}_{v_{\rm ext},\{N_I\}}[u,\{\Lambda_I\}]
=
E^{\rm v}_{v_{\rm ext}}[u]
+
\sum_I \Lambda_I
\left(
\rho^{\rm v}_I[u] - N_I
\right),
\end{equation}

where $\Lambda_I$ denotes the Lagrange multiplier associated with constraint $I$, while $N_I$
is the target value of the corresponding constrained quantity.  When constraining the local magnetization, $N_I$ represents the target value of the $I$th local magnetic moment, denoted by $\mathbf{M}_{I,\mathrm{target}}$.

For noncollinear magnetism, the electronic state is described through the spin-density matrix,

\begin{equation}
\rho_{ss'}({\bf r})
=
\sum_i f_i
\phi_{si}^*({\bf r})
\phi_{s'i}({\bf r}),
\end{equation}

with spin indices $s,s'\in\{\uparrow,\downarrow\}$.

Local magnetic moments are obtained from weighted integrations of the spin-density matrix
within atom-centered regions. For atom $\kappa$, the local magnetic moment is defined as

\begin{equation}                                                                   
\mathbf{M}_{\kappa} = \sum_{ss'}                                                   
\int d^3r \,w_{\kappa}\!\left(\mathbf r\right) \boldsymbol{\sigma}_{ss'}\,
\rho_{ss'}(\mathbf r),
\end{equation}

where $w_{\kappa}(\mathbf r)$ denotes the atom-centered weight function associated with atom $\kappa$
and $\boldsymbol{\sigma}$ are the Pauli matrices.

In Ref.~\onlinecite{Gonze2022}, it is shown how the Lagrange parameters $\Lambda_I$ can be obtained.
For simplicity, here we show the line of thought in the case of a single constraint, labeled $A$, with weight function
$w_A(\mathbf r)$. The projection of the potential residual onto the corresponding
constraint subspace is defined as
\begin{equation}
R_A^{\rm v}[u] =\int d\mathbf r\, R^{\rm v}[u](\mathbf r)\, w_A(\mathbf r),
\end{equation}
while the overlap of the weight function with itself is
\begin{equation}
W_{AA} =   \int d\mathbf r\, w_A(\mathbf r)w_A(\mathbf r).
\end{equation}
The corresponding Lagrange multiplier is
$\Lambda_A[u]=-R_A^{\rm v}[u](W_{AA})^{-1}$. Substituting this expression into the
augmented functional yields the potential-based cDFT functional
\begin{equation}
E^{\rm cDFT}_{v_{\rm ext},N_A}[u]
= E^{\rm v}_{v_{\rm ext}}[u] - R_A^{\rm v}[u](W_{AA})^{-1}
\left(\rho_A^{\rm v}[u]-N_A\right).
\end{equation}
This functional depends only on the trial screened potential $u$ and does not require
the Lagrange multiplier to be determined through a separate optimization procedure.
Consequently, the target constraint $N_A$ is treated as an external parameter on the
same footing as the atomic coordinates and lattice parameters.

The different gradients with respect to the external parameters, as well as to the constraints, can now be obtained.
The derivative of the total energy with respect to the constrained magnetic direction
defines the spin torque,

\begin{equation}
T_J=-\frac {\partial E^{\rm +v}_{v_{\rm ext},\{ \mathbf{M}_{J,\mathrm{target}}\}}[v^*]}
{\partial  \mathbf{M}_{J,\mathrm{target}}}
= \Lambda_J[v^*].
\label{eq:magnetic_torque}
\end{equation}

Owing to the stationary character of the cDFT functional at the self-consistent screened potential $v^*$, the Hellmann-Feynman 
theorem can be applied directly to constrained calculations. The force acting on
atom $\kappa$ along Cartesian direction $\alpha$ is therefore written as

\begin{eqnarray}
F_{\kappa\alpha} &=& \left.
-\frac {\partial E^{\rm v}_{v_{\rm ext}}}
{\partial \tau_{\kappa\alpha}} \right|_{v^*}+\sum_J T_J\left.
\frac {\partial \rho^{\rm v}_J[u]} {\partial \tau_{\kappa\alpha}} \right|_{v^*},
\end{eqnarray}

where $\tau_{\kappa\alpha}$ denotes the atomic coordinate.

Similarly, the stress tensor is obtained from the derivative of the total energy with respect to the strain tensor $\eta_{\alpha\beta}$,

\begin{equation}
%\sigma_{\alpha\beta}
%=
%\frac{1}{\Omega}
%\frac{\partial E^{\rm SC}_{\Omega{\rm tot}}}
%{\partial \eta_{\alpha\beta}},
\sigma_{\alpha\beta}= \frac{1}{\Omega} \left.
\frac   {\partial E^{\rm v}_{\Omega v_{\rm ext}}}
  {\partial \eta_{\alpha\beta}}\right|_{v^*}
-\sum_J T_J \left.\frac{1}{\Omega}
\frac   {\partial \rho^{\rm v}_J[u]}    {\partial \eta_{\alpha\beta}}
\right|_{v^*}
\end{equation}
where $\Omega$ is the unit-cell volume. The deformation tensor is defined through

\begin{equation}
r_{\alpha}^{\prime}=r_{\alpha}+\sum_{\beta}\eta_{\alpha\beta} r_{\beta}.
\end{equation}

Within the present implementation, atomic coordinates, lattice vectors, and local 
magnetic moments are optimized simultaneously during the transition-path calculations described below.

\section{Finding the Transition Path}
\label{Sec:method_NEB}

The present implementation is based on a generalized nudged elastic band (NEB) formalism 
in which atomic coordinates, lattice degrees of freedom, and magnetic variables 
are optimized simultaneously along the transition pathway.
The method unifies standard NEB,
 climbing-image NEB (CI-NEB), generalized solid-state NEB (G-SSNEB), and variable-cell NEB
  formulations within a single framework.
\subsection{Path Initialization}
For each image $i$, the generalized coordinate vector is defined as

\begin{equation}
	\mathbf{X}_i = 	\left\{\mathbf{R}_i,\mathbf{h}_i,\mathbf{M}_i 	\right\},
\end{equation}

where $\mathbf{R}_i$ denotes the atomic coordinates, $\mathbf{h}_i$ the lattice degrees 
of freedom, and $\mathbf{M}_i$ the magnetic degrees of freedom.

\begin{figure}
	\centering
	\includegraphics[width=0.45\textwidth]{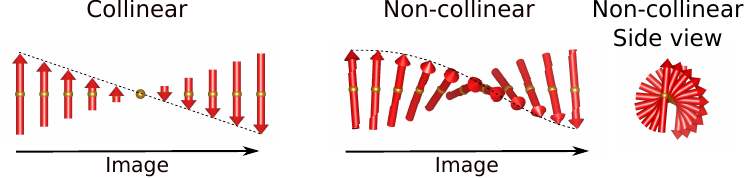}
\caption{
Schematic illustration of the interpolation of magnetic moments between adjacent images.
Collinear moments are interpolated linearly (left), while noncollinear spin orientations are 
interpolated using spherical linear interpolation (Slerp) along the shortest geodesic on the unit sphere (right).
}
	\label{fig:slerp}
\end{figure}

The initial pathway is constructed by interpolation between the initial and final images. Atomic 
coordinates, lattice vectors, and collinear magnetic moments are interpolated linearly. For noncollinear
magnetic configurations, a geometrically consistent interpolation of the spin orientations is required. 
Accordingly, the magnetic moment magnitudes are interpolated linearly, whereas their orientations are 
initialized using spherical linear interpolation (Slerp)\cite{Shoemake1985}, 
\begin{equation}
	\mathrm{Slerp}(\hat{\mathbf{M}}_1,\hat{\mathbf{M}}_2;u)=
	\frac{\sin[(1-u)\theta]}{\sin\theta}\hat{\mathbf{M}}_1
	+ 	\frac{\sin(u\theta)}{\sin\theta}\hat{\mathbf{M}}_2,
\end{equation}

where $\hat{\mathbf{M}}_1 \cdot \hat{\mathbf{M}}_2 = \cos\theta$ and $u \in [0,1]$. The spin orientations are thus 
interpolated along the shortest geodesic on the unit sphere, ensuring a smooth and geometrically 
consistent initialization of the magnetic pathway. The distinction between linear and spherical
interpolation is illustrated schematically in Fig.~\ref{fig:slerp}.     
Whereas linear interpolation is appropriate for collinear magnetic configurations, noncollinear spin 
orientations must be interpolated on the spin manifold to preserve their rotational character.
\subsection{Generalized Energy Gradients}

For each image, the energy derivatives are evaluated with respect to the atomic,
lattice, and magnetic degrees of freedom. At finite external pressure, the relevant
thermodynamic potential is the enthalpy
\begin{equation}
\mathcal{H}_i = E_i+P_{\rm ext}\Omega_i,
\end{equation}
which reduces to the total energy $E_i$ when $P_{\rm ext}=0$.

\subsubsection{Atomic Degrees of Freedom}

The force acting on atom $\kappa$ in image $i$ is defined as
\begin{equation}
\mathbf{F}^{\rm at}_{i\kappa} =- \frac{\partial\mathcal{H}_i}
{\partial\mathbf{R}_{i\kappa}},
\end{equation}
where $\mathbf{R}_{i\kappa}$ denotes the position of atom $\kappa$ in image $i$.

\subsubsection{Variable-Cell Degrees of Freedom}

To describe structural transformations involving strain and lattice deformation,
the generalized configurational space includes cell degrees of freedom following
the G-SSNEB formalism of Sheppard \textit{et al.}~\cite{Sheppard2012}.

The lattice deformation relative to a reference cell is written as
\begin{equation}
\Delta\mathbf h_i = \mathbf h_i-\mathbf h^{0},
\end{equation}
and the corresponding generalized lattice coordinate is defined as
\begin{equation}
\mathbf X_i^{\rm cell} =\eta_i \left(\mathbf h^{0}\right)^{-T}
\Delta\mathbf h_i,
\end{equation}
where
\begin{equation}
\eta_i = \Omega_i^{1/3}N^{1/6},
\end{equation}
with $\Omega_i$ denoting the cell volume and $N$ the number of atoms.
To keep the Jacobian constant throughout the NEB optimization, the image-dependent
volume $\Omega_i$ is replaced by the average volume of the fixed initial and final
images~\cite{Sheppard2012},
\begin{equation}
\overline{\Omega} = \frac{\Omega_{\rm initial}+\Omega_{\rm final}}{2}.
\end{equation}
The scaling factor used for all images is therefore
\begin{equation}
\eta = \overline{\Omega}^{1/3}N^{1/6}.
\end{equation}

The generalized lattice force is obtained from the stress tensor as
\begin{equation}
\mathbf F_i^{\rm cell} = - \frac{\Omega_i}{\eta} \boldsymbol{\sigma}_i,
\end{equation}
where
\begin{equation}
\boldsymbol{\sigma}_i = \boldsymbol{\sigma}_i^{\rm int}
+ P_{\rm ext}\mathbf I.
\end{equation}
Here, $\boldsymbol{\sigma}_i^{\rm int}$ is the internal stress tensor,
$P_{\rm ext}$ is the external hydrostatic pressure, and $\mathbf I$ is
the identity tensor.

\subsubsection{Magnetic Degrees of Freedom}

Magnetic degrees of freedom are incorporated into the generalized configurational
space whenever the magnetic configuration changes along the transition pathway.
For each image $i$, the magnetic contribution to the generalized force is given by
the cDFT magnetic torque $\mathbf T_i$ defined in
Eq.~\eqref{eq:magnetic_torque}. 
For noncollinear configurations, the local magnetic moments are updated according to the magnetic torque 
$\mathbf T_i$ , allowing both their orientations and magnitudes to evolve during the pathway optimization.

The atomic, lattice, and magnetic contributions are collected into the generalized
physical-force vector
\begin{equation}
\boldsymbol{\mathcal F}_i = \left\{ \mathbf F_i^{\rm at},\mathbf F_i^{\rm cell},
\mathbf T_i \right\}.
\end{equation}

\subsection{Generalized Nudged Elastic Band Method}

Starting from the initialized pathway, the generalized NEB method relaxes the intermediate
images toward a minimum-energy path in the combined atomic, lattice, and magnetic
configurational space.
The generalized NEB force acting on image $i$ is decomposed into a perpendicular physical 
force and a parallel spring contribution,

\begin{equation}
\boldsymbol{\mathcal{F}}^{\mathrm{NEB}}_i = \boldsymbol{\mathcal{F}}^{\perp}_i
+ \boldsymbol{\mathcal{F}}^{\mathrm{spring}}_i,
\end{equation}

The perpendicular component is obtained by projecting out the component parallel to the tangent direction,

\begin{equation}
\boldsymbol{\mathcal{F}}^{\perp}_i
= \boldsymbol{\mathcal{F}}_i - \left(
\boldsymbol{\mathcal{F}}_i\cdot\hat{\boldsymbol{\tau}}_i
\right) \hat{\boldsymbol{\tau}}_i.
\end{equation}

where $\hat{\tau}_i$ denotes the normalized tangent vector.

The tangent construction follows the improved prescription of Henkelman and Jónsson.~\cite{Henkelman2000b} Defining

\begin{equation}
	\Delta \mathbf{X}_i =\mathbf{X}_i-\mathbf{X}_{i-1},
\end{equation}

the tangent vector is evaluated according to the local energy ordering,

\begin{equation}
	\tau_i= 
    \begin{cases}
		\Delta \mathbf{X}_{i+1},& E_{i+1}>E_i>E_{i-1},
		\\[6pt]
		\Delta \mathbf{X}_{i}, & E_{i+1}<E_i<E_{i-1}.
	\end{cases}
\end{equation}

At extrema along the pathway, an energy-weighted interpolation is used,

\begin{equation}
	\tau_i=
	\begin{cases}
		\Delta \mathbf{X}_{i+1}\Delta E_{\max}
		+\Delta \mathbf{X}_{i}\Delta E_{\min},
		& E_{i+1}>E_{i-1}, 	\\[6pt]
		\Delta \mathbf{X}_{i+1}\Delta E_{\min}
		+ \Delta \mathbf{X}_{i}\Delta E_{\max}, 
        &E_{i+1}<E_{i-1}.
	\end{cases}
\end{equation}

where

\begin{equation}
	\Delta E_{\max} =\max 	\left(	|E_{i+1}-E_i|,
	|E_{i-1}-E_i|\right),
\end{equation}

and

\begin{equation}
	\Delta E_{\min} =\min\left(|E_{i+1}-E_i|,|E_{i-1}-E_i|	\right).
\end{equation}

The normalized tangent vector is then

\begin{equation}
	\hat{\tau}_i=\frac{\tau_i}{|\tau_i|}.
\end{equation}

The spring force is evaluated using the distance-based formulation,

\begin{equation}
	\mathcal{F}^{\mathrm{spring}}_i =\left[
	k_{i+1}|\mathbf{X}_{i+1}-\mathbf{X}_i|
	-k_i  |\mathbf{X}_i-\mathbf{X}_{i-1}|
	\right] \hat{\tau}_i,
\end{equation}

where $k_i$ denotes the spring constant associated with image $i$.

Both constant and adaptive spring schemes are implemented. In the adaptive formulation, 
the spring constants increase near the saddle-point region in order to improve the image
 distribution around the transition state.

\subsection{Climbing-Image NEB}

The implementation includes the climbing-image extension of the NEB method. After an initial
 relaxation stage, the image with the highest energy is converted into a climbing image by
  removing the spring-force contribution and inverting the parallel component of the physical force,

\begin{equation}
	\mathcal{F}^{\mathrm{CI}}_i =
	\mathcal{F}_i -	2\left( \mathcal{F}_i \cdot \hat{\tau}_i
	\right)  \hat{\tau}_i.
\end{equation}

This procedure drives the climbing image toward the first-order saddle point.

\subsection{Numerical Optimization}

The generalized NEB forces are relaxed using several optimization algorithms,
including steepest descent (SD)~\cite{nocedal2006}, local limited-memory
Broyden-Fletcher-Goldfarb-Shanno (LL-BFGS), and global limited-memory BFGS
(GL-BFGS)~\cite{Sheppard2008}. Both BFGS schemes retain only a limited history
of coordinate and force differences. The distinction between them concerns the
configuration space over which the approximate inverse Hessian is constructed.

In the steepest-descent method, the generalized coordinates of each image are
updated along the corresponding NEB force,
\begin{equation}
\mathbf X_i^{(n+1)}
=
\mathbf X_i^{(n)}
+
\alpha\,
\boldsymbol{\mathcal F}_i^{\rm NEB,(n)},
\end{equation}
where $\alpha$ is the optimization step size. In practice, separate step-size
parameters are used for the structural and magnetic degrees of freedom to account
for their different energy scales and convergence behavior. The atomic and lattice
variables are updated using $\alpha_{\rm str}$, whereas the magnetic variables are
updated using $\alpha_{\rm mag}$.

In the LL-BFGS scheme, each intermediate image has an independent optimization
history and an independent approximate inverse Hessian. The search direction for
image $i$ is therefore evaluated as
\begin{equation}
\mathbf d_i^{(n)}
=
\left(\mathbf H_{i,n}^{\rm LL}\right)^{-1}
\boldsymbol{\mathcal F}_i^{\rm NEB,(n)},
\end{equation}
and the image is updated according to
\begin{equation}
\mathbf X_i^{(n+1)}
=
\mathbf X_i^{(n)}
+
\lambda_{i,n}\mathbf d_i^{(n)}.
\end{equation}
Although all images are updated during the same NEB iteration, the inverse-Hessian
approximations are constructed image by image. Consequently, the LL-BFGS
approximation does not explicitly contain the off-diagonal Hessian blocks describing
couplings between neighboring images.

In the GL-BFGS scheme, all $P$ intermediate images are treated collectively.
The band-wide coordinate vector $\mathbf X_{\rm band}^{(n)}$ is formed by
concatenating the generalized coordinates $\mathbf X_i^{(n)}$ of all images, while
the corresponding force vector $\boldsymbol{\mathcal F}_{\rm band}^{(n)}$ is formed
by concatenating their generalized NEB forces
$\boldsymbol{\mathcal F}_i^{\rm NEB,(n)}$. The entire band is then updated using a
single approximate inverse Hessian,
\begin{equation}
\mathbf X_{\rm band}^{(n+1)}
=
\mathbf X_{\rm band}^{(n)}
+
\lambda_n
\left(\mathbf H_n^{\rm GL}\right)^{-1}
\boldsymbol{\mathcal F}_{\rm band}^{(n)}.
\end{equation}
Unlike LL-BFGS, which constructs an independent inverse-Hessian approximation for
each image, GL-BFGS includes inter-image couplings by treating the complete band
within a single optimization space.
 
	%%%%%%%%%%%%%%%%%%%%%%%%%%
	\section{Results and discussion}
	%\section{3. Results and discussion}
	\label{Sec:results_discussion}
	%%%%%%%%%%%%%%%%%%%%%%%%%%

To validate the proposed generalized cDFT-NEB framework, we consider a series of 
representative magnetic and magnetostructural transitions. These examples are designed
to assess the capability of the method to simultaneously treat atomic, lattice, 
and magnetic degrees of freedom within the potential-based constrained 
density-functional theory (cDFT) formalism.

We first investigate the first-order metamagnetic transition in FeRh between 
the antiferromagnetic (AFM) and ferromagnetic (FM) phases by comparing 
fixed-cell and variable-cell transition pathways. 
We then examine the more challenging direct magnetostructural transformation
from bcc-FM to hcp-NM iron, involving substantial lattice deformation, atomic
shuffle, volume reduction, and collapse of the local magnetic moments.
Finally, we study the collective spin reversal in monolayer CrI$_3$, 
in which the transition is governed predominantly by the evolution of the magnetic 
moments while the accompanying atomic and lattice relaxations remain negligible. 
This system also provides a convenient platform for assessing the influence 
of spin-orbit coupling (SOC) on the minimum-energy path and the associated magnetic anisotropy barrier.

Together, these applications demonstrate the broad applicability of the proposed framework, 
spanning strongly coupled magnetostructural transformations, 
predominantly magnetic transitions, and systems involving noncollinear spin rotations.

	\subsection{Computational Details}

All calculations were performed within the framework of density-functional theory (DFT) as implemented in the 
ABINIT software  (version 10.8)~\cite{Gonze2020,Romero2020,Verstraete2025}. 
The electron-ion interactions  for Fe and FeRh were described using the projector augmented-wave (PAW) 
method~\cite{TORRENT2008337} with datasets obtained from 
the \texttt{Pseudo-Dojo} project~\cite{VanSetten2018}. To accurately represent 
the rapidly varying density and potential terms within the PAW augmentation spheres, an auxiliary double-grid energy cutoff of $80\text{ Ha}$ was 
employed. Conversely, optimized norm-conserving pseudopotentials were employed for $\mathrm{CrI_3}$. Exchange-correlation effects were treated within 
the Perdew-Burke-Ernzerhof (PBE) generalized-gradient approximation (GGA)~\cite{perdew1996} for Fe and FeRh, while the local-density 
approximation (LDA)\cite{perdew1992} was adopted for $\mathrm{CrI_3}$.
Spin-orbit coupling was included in all calculations.

For Fe and FeRh, the Kohn-Sham wave functions were expanded in a plane-wave 
basis set with a kinetic-energy cutoff of $25\text{ Ha}$.
Brillouin-zone integrations employed Monkhorst-Pack $k$-point meshes of $16\times16\times10$ 
for the two-atom Fe cells and $8\times8\times8$ for the four-atom FeRh cells. 
Electronic occupations were treated using a Gaussian smearing scheme with a 
broadening width of $10^{-2}\text{ Ha}$ to facilitate self-consistent field 
convergence. For the monolayer $\mathrm{CrI_3}$ system, a higher plane-wave 
kinetic-energy cutoff of $50\text{ Ha}$ and an $8\times8\times1$ 
Monkhorst-Pack $k$-point mesh were required. The $\mathrm{CrI_3}$ simulation 
cell contained eight atoms and included a vacuum spacing of approximately 
$18.5\text{ \AA}$ along the out-of-plane direction to eliminate spurious 
interactions between periodic images.

Ground-state structures were fully relaxed until the residual atomic forces 
fell below $10^{-4}\text{ Ha/Bohr}$. Noncollinear magnetic configurations 
were treated explicitly using spinor wave functions within the 
spin-density-matrix formalism. Unless otherwise specified, crystal symmetry 
operations were disabled throughout the nudged elastic band (NEB) 
calculations to permit the unrestricted evolution of the atomic, lattice, 
and magnetic degrees of freedom.

Minimum-energy pathways (MEPs) were determined using the generalized 
cDFT-NEB implementation described in Sec.~\ref{Sec:method_NEB}, employing the 
improved-tangent formulation of Henkelman and 
J{\'o}nsson~\cite{Henkelman2000b}. For fixed-cell calculations, 
steepest-descent, LL-BFGS, and GL-BFGS optimizers were benchmarked and 
found to converge to identical MEPs and activation barriers within numerical 
accuracy. Variable-cell calculations were executed via the SD optimizer, 
enabling the simultaneous relaxation of atomic coordinates, lattice vectors, 
and magnetic degrees of freedom within the potential-based constrained DFT 
framework.
Unless stated otherwise, the NEB calculations were considered converged when
the mean, over all dynamically optimized images, of the absolute total-energy
difference between two successive image iterations fell below $10^{-5}$ Ha.

\subsection{FeRh: AFM-FM Metamagnetic Transition}

Equiatomic FeRh is a prototypical metamagnetic alloy exhibiting a first-order transition 
from an antiferromagnetic (AFM) state to a ferromagnetic (FM) state at an experimental 
transition temperature of approximately $340$-$370$ K\cite{Kouvel1962,Shirane1964,Annaorazov1992}. 
 The material crystallizes in the chemically ordered B2 (CsCl-type) structure with 
 cubic space group $Pm\bar{3}m$ (No.~221), where Fe atoms occupy the cube corners 
 and Rh atoms occupy the body-centered positions. 
 The AFM-FM transition is isostructural and is accompanied by a small isotropic
 lattice expansion, making FeRh a benchmark system for studying coupled magnetic and lattice degrees of freedom.

Because the transition occurs close to room temperature and can be manipulated 
by temperature, strain, magnetic field, and ultrafast excitation, FeRh has 
attracted considerable attention for spintronic and magnetocaloric applications~\cite{Lewis2016,Annaorazov1992}. 
Owing to its well-established magnetic transition pathway and the availability
of previous string-method and DeltaSpin calculations \cite{Tang2025}, 
 FeRh provides an ideal test system for validating the present generalized NEB implementation.

In the low-temperature AFM phase, Fe magnetic moments are aligned antiferromagnetically 
while the Rh moments vanish by symmetry. In the FM phase, the Fe moments align
parallel and finite induced magnetic moments develop on the Rh atoms. In the 
present calculations, the AFM configuration was taken as the initial state and the FM
configuration as the final state. A four-atom cell corresponding to the primitive AFM FeRh cell was employed.

To facilitate comparison with previous magnetic string-method calculations
\cite{Tang2025}, the transition pathway was discretized using 11 images. The initial 
magnetic path was generated by prescribing one Fe moment to remain aligned along the
positive spin direction while the second Fe moment gradually rotates from antiparallel
to parallel alignment. A middle image was constructed with the two Fe moments perpendicular
to each other while preserving approximately constant magnetic-moment amplitudes. 
The remaining images were initialized using spherical interpolation (Slerp) between
neighboring magnetic configurations, ensuring smooth rotational evolution of the local 
moments during the NEB optimization.

Both fixed-cell and variable-cell calculations were performed. For the fixed-cell 
calculations, the lattice parameters were constrained to the equilibrium AFM structure,
whereas in the variable-cell calculations both lattice and magnetic degrees of freedom 
were optimized simultaneously along the pathway. The generalized NEB calculations employed 
the improved-tangent formalism with 11 images.
Three optimization schemes were tested for the fixed-cell calculations, namely 
steepest descent (SD), LL-BFGS, and  GL-BFGS optimizations. All three methods converged to 
identical minimum-energy pathways and barrier heights, demonstrating the robustness of the 
present implementation. In the current implementation, variable-cell calculations were 
performed using the steepest-descent optimizer.

\begin{figure}
	\centering
	\includegraphics[width=0.45\textwidth]{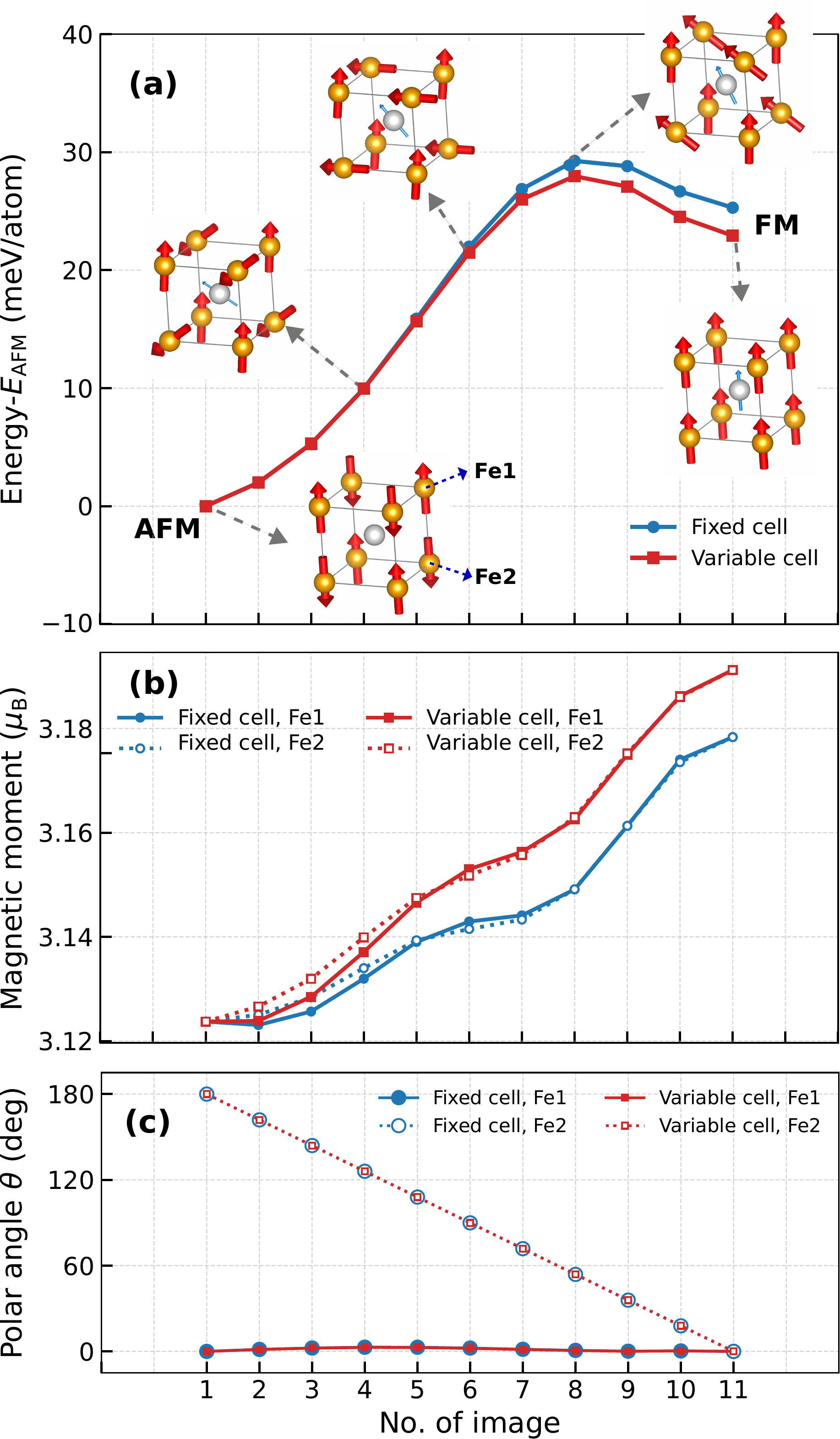}
	\caption{{%\color{red}
     Antiferromagnetic (AFM) to ferromagnetic (FM) transition in FeRh 
    obtained using the generalized cDFT-NEB method.
    (a) Minimum-energy pathways obtained from fixed-cell (blue circles)
    and variable-cell (red squares) calculations.
    Representative magnetic configurations are illustrated for selected
    images along the pathways. Fe and Rh atoms are shown in gold and gray,
    respectively, while the arrows indicate the local magnetic moments.
    Fe1 denotes the Fe site whose magnetic moment remains approximately
    aligned along the positive $z$ direction, whereas Fe2 denotes the
    Fe site whose magnetic moment rotates from antiparallel to parallel
    alignment with Fe1.
    The transition proceeds through intermediate noncollinear magnetic
    configurations. Allowing lattice relaxation slightly reduces the
    energy barrier, highlighting the role of magnetovolume coupling.
    (b) Evolution of the magnetic moment magnitudes
    $|\mathbf{M}_1|$ and $|\mathbf{M}_2|$ of Fe1 and Fe2, respectively,
    along the fixed-cell and variable-cell pathways.
  (c) Evolution of the polar angles $\theta$ of the Fe1 and Fe2
magnetic moments with respect to the positive $z$ axis along
the fixed-cell and variable-cell pathways. The angle for Fe1
remains close to $0^\circ$, with small deviations at intermediate
images, whereas that for Fe2 decreases from $180^\circ$ to $0^\circ$.}
	}
	\label{fig:FeRh_path}
\end{figure}
The resulting minimum-energy pathways are presented in Fig.~\ref{fig:FeRh_path}. 
The magnetic transition proceeds continuously through noncollinear intermediate states connecting
the AFM and FM configurations. 
{%\color{red}
During the transition, the Fe magnetic moments undergo a gradual reorientation
accompanied by a small increase in their magnitudes, while finite induced magnetic moments 
develop on the Rh atoms.

As shown in Fig.~\ref{fig:FeRh_path}(b), the magnitudes of the Fe magnetic moments increase from approximately
$3.124\,\mu_B$ in the AFM state to $3.178\,\mu_B$ and $3.191\,\mu_B$ in the FM state for the fixed-cell and
variable-cell calculations, respectively. Small differences between the magnitudes of the Fe1 and Fe2 moments are
observed at intermediate images in the calculations including spin--orbit coupling (SOC). The induced Rh
magnetic moments increase from zero in the AFM state to $1.056\,\mu_B$ and $1.054\,\mu_B$ in the FM state for the
fixed-cell and variable-cell calculations, respectively.

Figure~\ref{fig:FeRh_path}(c) shows the evolution of the polar angles of the Fe magnetic moments with respect to
the positive $z$ axis. The Fe1 moment remains nearly aligned with the positive $z$ direction, exhibiting small
angular deviations at intermediate images, whereas the polar angle of Fe2 decreases continuously from
$180^\circ$ to $0^\circ$, corresponding to its rotation from antiparallel to parallel alignment with Fe1.
The magnetic moments rotate predominantly within the $xz$ plane, with negligible components along the
$y$ direction.
}

For the fixed-cell and variable-cell calculations, the energy barrier is found to be approximately $29$ meV/atom 
and $28$ meV/atom, respectively, in good agreement with previous DeltaSpin and magnetic string-method studies
\cite{Tang2025}
reporting barriers of $\sim$ 33-35 meV/atom. Using the relation $E \approx k_B T$, this barrier
corresponds to a characteristic temperature close to the experimentally observed AFM-FM transition temperature range.

Allowing simultaneous lattice relaxation lowers the energy barrier further, highlighting 
the importance of magnetovolume coupling in FeRh. The variable-cell pathway exhibits a smoother
evolution of the total energy and a reduced saddle-point energy relative to the fixed-cell calculations,
consistent with the experimentally observed lattice expansion accompanying the metamagnetic transition.
These results demonstrate that the present generalized cDFT-NEB framework can naturally describe coupled
spin and lattice transitions within a unified formalism.
%%%%%%%%%%%%%%%%%%%%%%%%%%%%%%%%%%%%%%%%%%%%%%%%%%%%%%%%%%%%%%

\subsection{bcc-to-hcp structural and magnetic transition in iron}

As a more demanding benchmark of the generalized NEB implementation, we
consider elemental iron, a prototypical system in which structural stability
and magnetism are strongly coupled. In contrast to FeRh, where the crystal
structure remains essentially unchanged during the magnetic transition, the
bcc-to-hcp transformation in iron involves simultaneous changes in the
internal atomic coordinates, lattice degrees of freedom, and local magnetic
moments.

At ambient conditions, ferromagnetic body-centered cubic iron
(bcc-FM, $\alpha$-Fe) is the ground-state phase. Under pressure, iron
undergoes a martensitic transformation to the hexagonal close-packed phase,
accompanied by a reduction in volume and magnetization. Previous
first-principles work described the pressure-dependent transition pathway in
terms of strongly coupled shear and shuffle modes~\cite{Liu2009}, while
in situ x-ray tomography and diffraction measurements confirmed that the
transformation follows the Burgers orientation
relationship~\cite{Boulard2020}. Within the present computational setup,
the nonmagnetic hcp phase lies approximately $54$ meV/atom above the bcc-FM
ground state at zero pressure.

\begin{figure}[t]
\centering
\includegraphics[width=0.72\columnwidth]{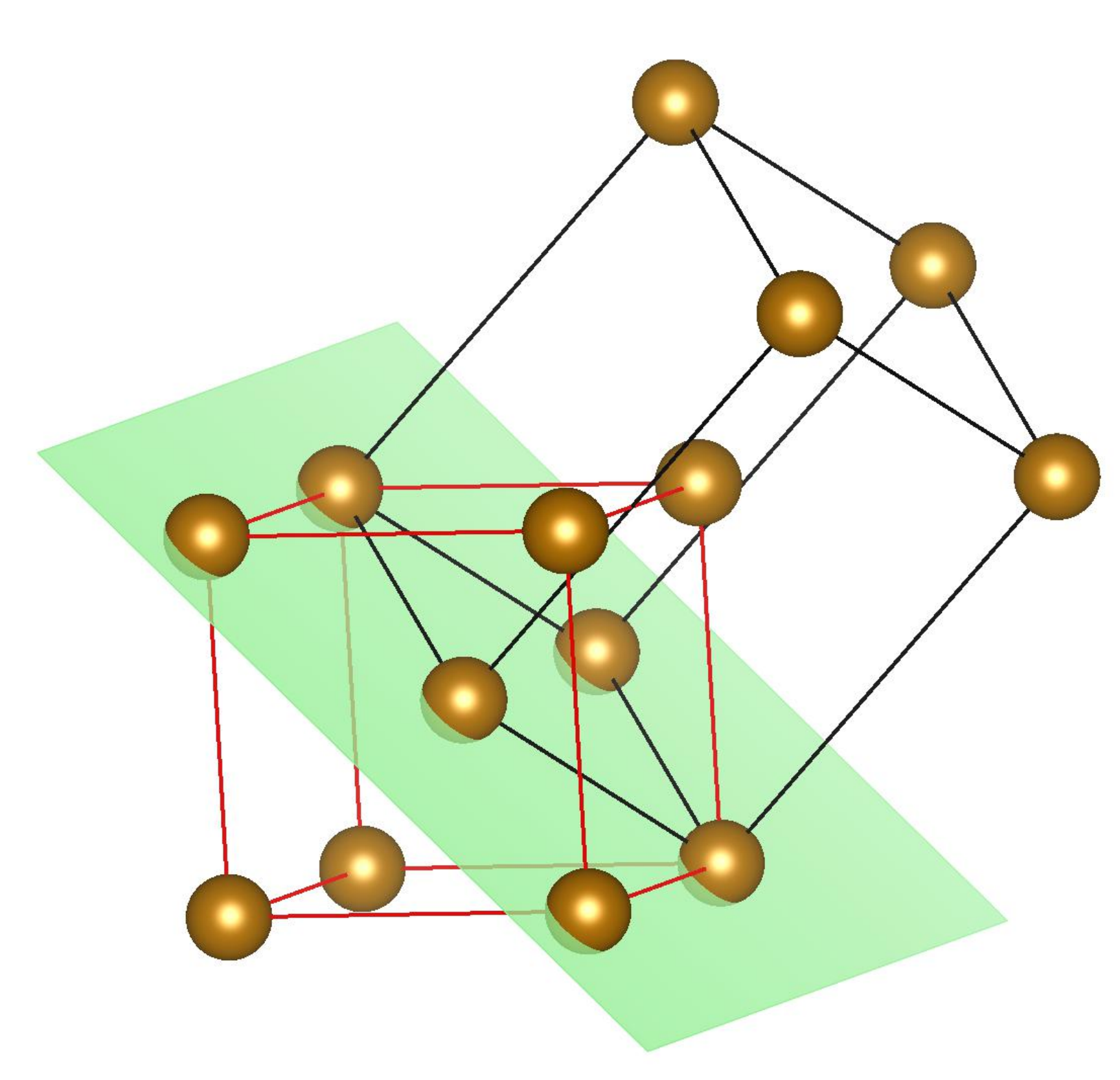}
\caption{
Construction of the reoriented bcc cell used for the bcc-to-hcp NEB
calculations. The conventional cubic bcc cell is outlined in red , while
the reoriented two-atom cell is outlined in black. The cell is selected such
that $(001)_{\mathrm{bcc}^{\prime}}$ corresponds to
$(110)_{\mathrm{bcc}}$ and evolves continuously into the
$(0001)_{\mathrm{hcp}}$ basal plane.
}
\label{fig:Fe_rotated_cell}
\end{figure}

For a continuous mapping between the bcc and hcp endpoints, the conventional
cubic representation of bcc Fe is replaced by the reoriented two-atom cell
illustrated in Fig.~\ref{fig:Fe_rotated_cell}. The cell is chosen such that
\begin{equation}
(001)_{\mathrm{bcc}^{\prime}}
\parallel
(110)_{\mathrm{bcc}}
\parallel
(0001)_{\mathrm{hcp}},
\end{equation}
where $\mathrm{bcc}^{\prime}$ denotes the reoriented representation used in
the NEB calculations. This change does not modify the physical bcc structure,
but provides a common two-atom periodicity for the two endpoints and
establishes a one-to-one correspondence between their atomic positions. The
transformation follows a Burgers-type mechanism in which a lattice
deformation converts the $(110)_{\mathrm{bcc}}$ planes into hexagonal nets,
while alternating planes
undergo a shuffle to generate the hcp stacking
sequence~\cite{Liu2009,Boulard2020}.

At zero pressure, the volume of the two-atom cell decreases from
$22.707$~\AA$^3$ in bcc-FM to $20.399$~\AA$^3$ in hcp-NM, corresponding
to a volume reduction of approximately $10.2\%$. The magnitude of the Fe
local magnetic moment decreases from approximately
$2.26\,\mu_{\mathrm{B}}$ in bcc-FM to zero in hcp-NM.

The transition pathway was represented by 15 images, including the two fixed
endpoints. To examine the influence of the magnetic initialization, two
different initial pathways were considered. In the first, the two Fe moments
were kept parallel in the initial band, while their magnitudes were
interpolated linearly from the bcc-FM value to zero at the hcp-NM endpoint.
This FM-like initialization resulted in a higher activation barrier and did
not reveal an intermediate metastable configuration.

In the second, AFM-assisted initialization, the relative orientation of the
two Fe moments was varied smoothly from parallel toward antiparallel
alignment. A perpendicular intermediate direction was specified
to define the rotation plane, and the intermediate orientations were
generated using spherical linear interpolation. Concurrently, the magnitudes
of both moments were progressively reduced to zero toward the hcp endpoint.
Starting from this band, the optimization converged to the lower-barrier
pathway.

\begin{figure}[h!]
\centering
\includegraphics[width=0.9\columnwidth]{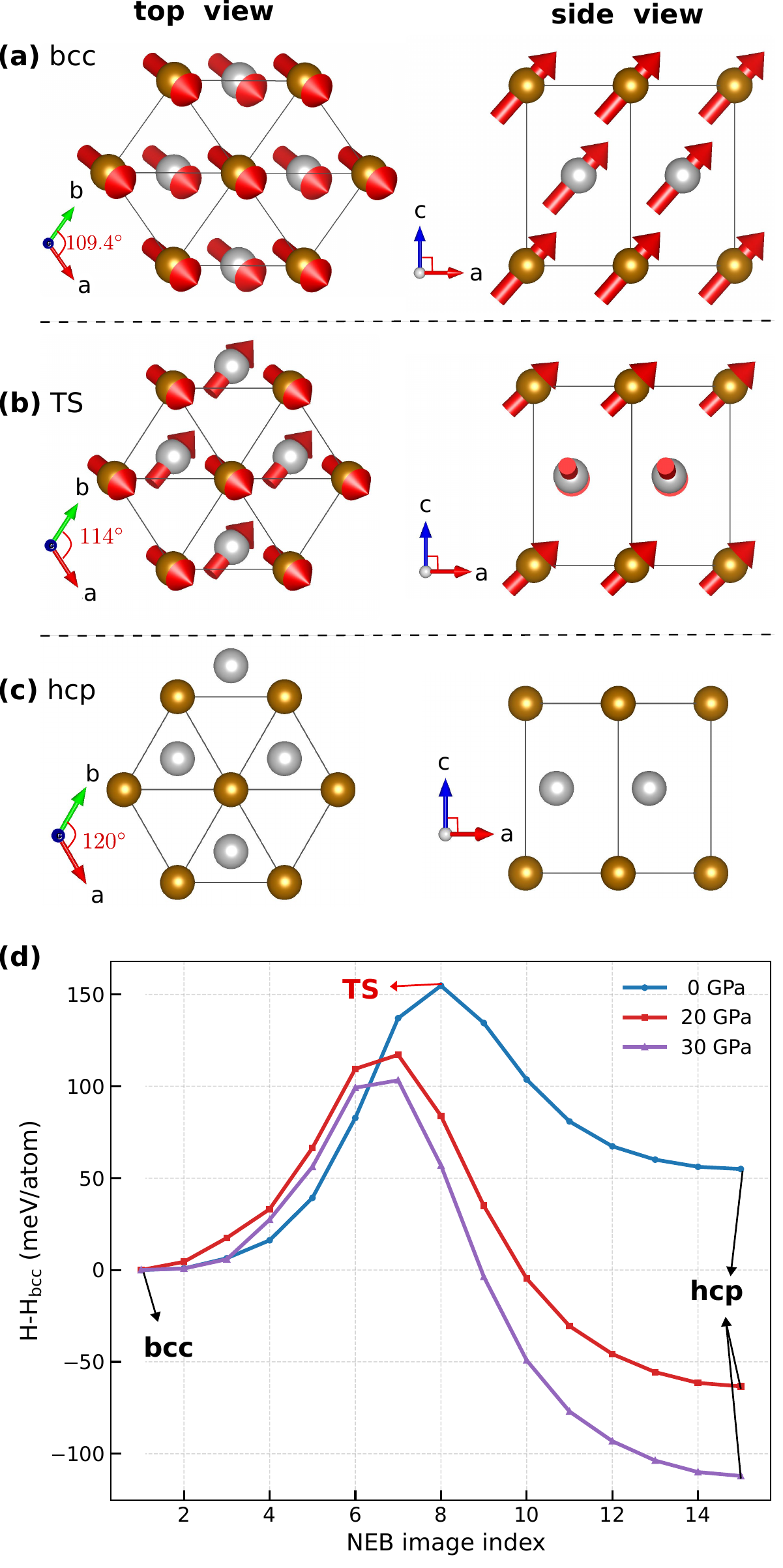}
\caption{
Top and side views of (a) the bcc-FM structure, (b) the zero-pressure
transition-state configuration, and (c) the hcp-NM structure. The gold and
gray spheres both represent Fe atoms, with the different colors used to
distinguish the two alternating atomic planes. During the transformation,
the gray Fe atom moves from the center of the
$(001)_{\mathrm{bcc}^{\prime}}$ plane, with reduced coordinates
$(1/2,1/2,1/2)$, toward the hcp position
$(1/3,2/3,1/2)$, or its symmetry-equivalent position, through the atomic
shuffle. The arrows indicate the local Fe magnetic moments. The in-plane
lattice angles are indicated in panels (a)-(c), while the corresponding
$c/a$ ratios are $1.633$, $1.645$, and $1.583$, respectively.
Panel (d) shows the enthalpy relative to the bcc-FM endpoint along the
15-image generalized NEB pathways at $0$, $20$, and $30$ GPa. The
corresponding activation barriers are $154$, $117$, and $103$ meV/atom,
respectively.
}
\label{fig:Fe_bcc_hcp}
\end{figure}

Under an imposed hydrostatic pressure, the relevant thermodynamic potential
is the enthalpy,
\begin{equation}
\mathcal{H}=E+P\Omega.
\end{equation}
The lower-barrier pathways obtained at $0$, $20$, and $30$ GPa are shown in
Fig.~\ref{fig:Fe_bcc_hcp}, where the enthalpy of each image is reported
relative to the bcc-FM endpoint at the same pressure. At zero pressure, the path
exhibits a single maximum with an activation barrier of approximately
$154$ meV/atom. Increasing pressure stabilizes the hcp phase relative to bcc
and lowers the activation barrier to approximately $117$ meV/atom at
$20$ GPa and $103$ meV/atom at $30$ GPa. No intermediate metastable minimum
is observed along these direct bcc-to-hcp pathways.

\begin{figure}[t]
\centering
\includegraphics[width=0.92\columnwidth]{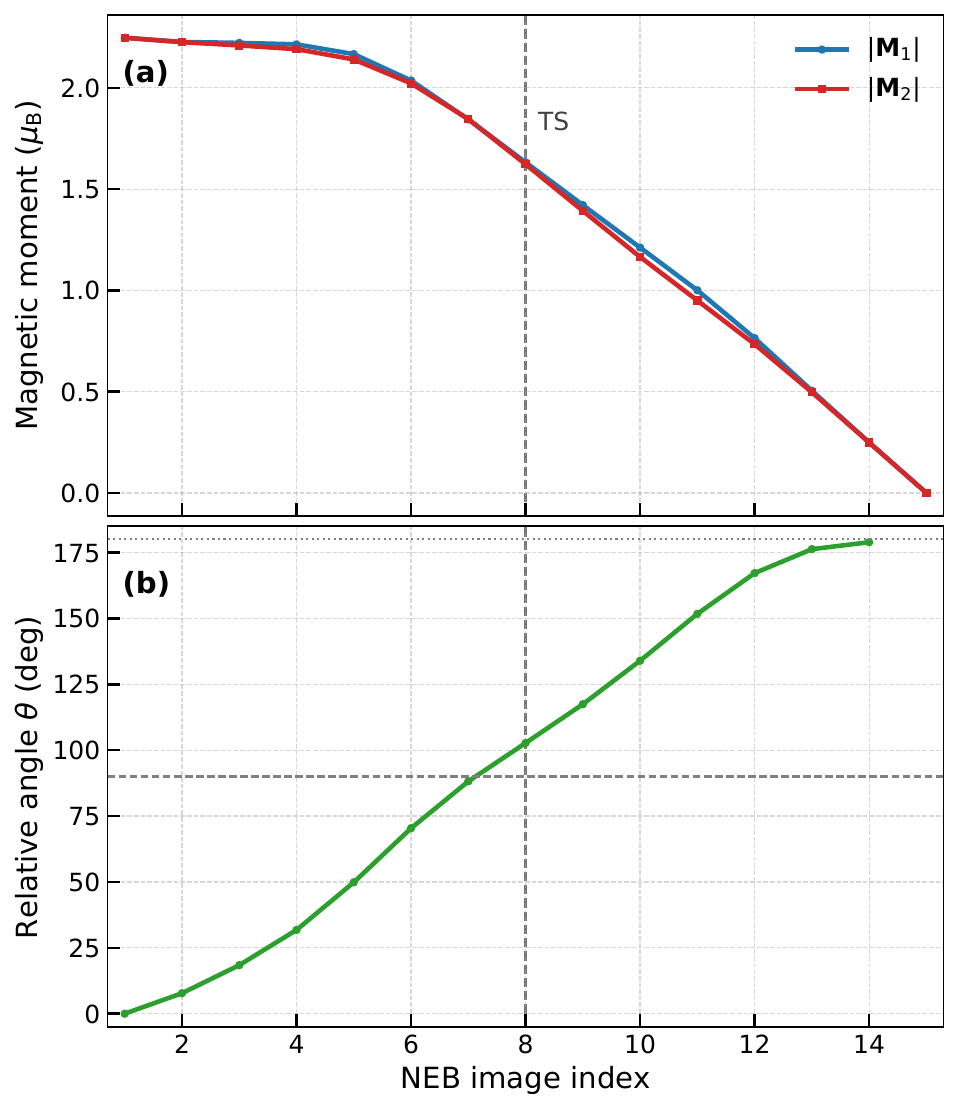}
\caption{
Evolution of the magnetic degrees of freedom along the zero-pressure
lower-barrier bcc-to-hcp pathway. The upper panel shows the magnitudes
$|\mathbf{M}_1|$ and $|\mathbf{M}_2|$ of the two Fe magnetic-moment vectors,
and the lower panel shows their relative angle $\theta$. The horizontal
dashed and dotted lines indicate $90^{\circ}$ and $180^{\circ}$,
respectively, and the vertical dashed line marks the transition state at
image 8. The angle is not reported for the final hcp-NM image because
both moments vanish and their relative orientation is therefore undefined.
Lines connecting the images are guides to the eye.
}
\label{fig:Fe_magnetic_path}
\end{figure}

The evolution of the magnetic degrees of freedom along the converged
zero-pressure lower-barrier pathway is resolved in
Fig.~\ref{fig:Fe_magnetic_path}. 
No symmetry constraint was imposed to enforce equal magnitudes of the two Fe magnetic moments. Nevertheless, $|\mathbf{M}_1|$ and $|\mathbf{M}_2|$ remain nearly equal along the optimized pathway 
and decrease from approximately $2.25\,\mu_{\mathrm{B}}$ at the bcc
endpoint to zero at the hcp endpoint. At the same time, their relative angle
increases continuously from $0^{\circ}$, passes close to $90^{\circ}$ at the
seventh image, reaches approximately $103^{\circ}$ at the transition state,
and approaches $180^{\circ}$ before the moments collapse. Thus, the magnetic
part of the pathway involves both a progressive rotation and a reduction of
the local-moment magnitudes, rather than a purely longitudinal collapse.

\begin{figure}[t]
\centering
\includegraphics[width=0.92\columnwidth]{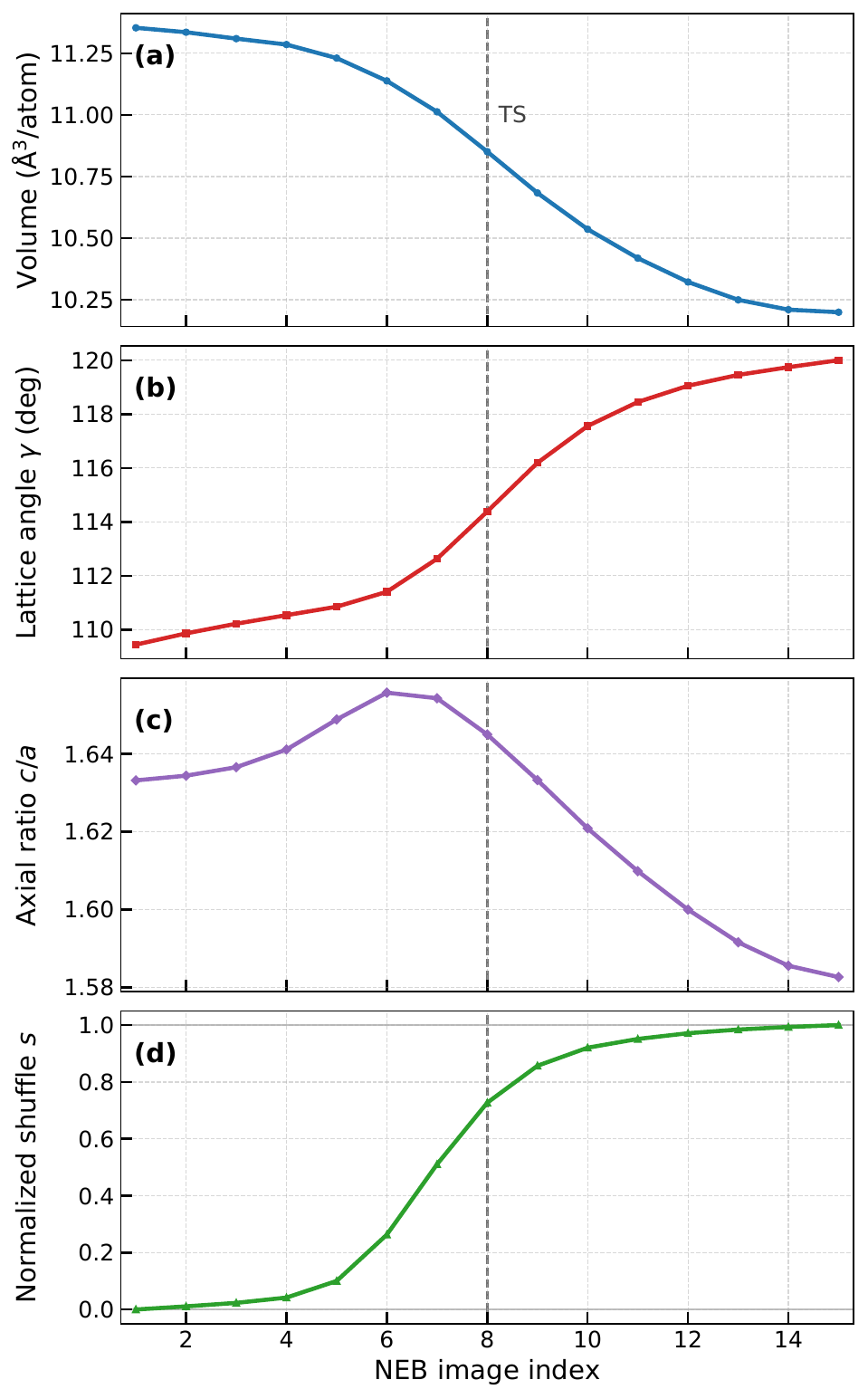}
\caption{
Structural evolution along the zero-pressure lower-barrier bcc-to-hcp
pathway. (a) Volume per atom, (b) in-plane lattice angle $\gamma$,
(c) axial ratio $c/a$, and (d) normalized atomic-shuffle coordinate $s$.
For the slightly distorted intermediate cells, $a$ in panel (c) is defined
as the mean length of the first two in-plane lattice vectors. The shuffle
coordinate is the projection of the second Fe reduced coordinate onto the
segment connecting $(1/2,1/2,1/2)$ in bcc to $(1/3,2/3,1/2)$ in hcp, so
that $s=0$ for bcc and $s=1$ for hcp. The vertical dashed line marks the
zero-pressure transition state at image 8. Lines connecting the images are
guides to the eye.
}
\label{fig:Fe_structural_path}
\end{figure}

The structural evolution along the zero-pressure pathway is
quantified in Fig.~\ref{fig:Fe_structural_path}. In addition to the monotonic
volume reduction, the in-plane lattice angle $\gamma$ increases smoothly
from $109.44^{\circ}$ in bcc to $120^{\circ}$ in hcp. The axial ratio $c/a$
shows a weak nonmonotonic variation: it increases from $1.633$ at the bcc
endpoint to a maximum of approximately $1.656$ near the sixth image and then
decreases to $1.583$ in hcp. At the transition state (image 8), the volume is
$10.851$~\AA$^3$/atom, $\gamma=114.38^{\circ}$, and $c/a=1.645$.

To separate the internal atomic displacement from the homogeneous cell
deformation, we define a normalized shuffle coordinate by projecting the
reduced coordinate $\mathbf{u}_i$ of the second Fe atom onto the line joining
its bcc and hcp endpoint positions,
\begin{equation}
s_i =
\frac{
(\mathbf{u}_i-\mathbf{u}_{\mathrm{bcc}})\!\cdot\!
(\mathbf{u}_{\mathrm{hcp}}-\mathbf{u}_{\mathrm{bcc}})
}{
|\mathbf{u}_{\mathrm{hcp}}-\mathbf{u}_{\mathrm{bcc}}|^2
},
\end{equation}
where $\mathbf{u}_{\mathrm{bcc}}=(1/2,1/2,1/2)$ and
$\mathbf{u}_{\mathrm{hcp}}=(1/3,2/3,1/2)$. With this definition, $s=0$ and
$s=1$ correspond to the bcc and hcp endpoints, respectively. The shuffle
remains small in the first part of the pathway, increases rapidly through the
central images, and reaches $s=0.727$ at the transition state. The correlated
changes in $\gamma$, $c/a$, and $s$ demonstrate that the shear, axial
distortion, and atomic shuffle are strongly coupled along the transformation
pathway.

The bcc-to-hcp transformation therefore cannot be described solely as a
lattice deformation. It combines a substantial volume reduction, coupled
cell deformation and atomic shuffle, and the progressive rotation and
collapse of the local magnetic moments. The lower barrier reached from the
AFM-assisted initialization further highlights the importance of treating
magnetic degrees of freedom explicitly when determining transition pathways
in iron.

\subsection{Spin reversal in monolayer CrI$_3$}

As a third application, we consider the collective magnetization reversal in a
ferromagnetic monolayer of CrI$_3$. In the initial state, the Cr magnetic moments
are aligned along the out-of-plane $+z$ direction, while the small induced moments
on the I atoms are oriented antiparallel to them. In the final state, all these
moments are reversed. Spin-orbit coupling is included in the calculations, giving
rise to a finite magnetic anisotropy barrier between the two magnetic states. The
transition pathway is represented by 13 images, including the initial and final
states.

In contrast to the FeRh and Fe transformations discussed above, the spin-reversal
process in CrI$_3$ induces only negligible atomic and lattice relaxations.
Nevertheless, the coupled magnetostructural framework is retained throughout the
pathway optimization, allowing structural and magnetic degrees of freedom to
evolve self-consistently. This example therefore probes the limit in which the
transition is governed predominantly by magnetic degrees of freedom. The pathway
optimization was performed using an energy-convergence threshold of
$2.0\times10^{-6}$ Ha.

\begin{figure}[t]
\centering
\includegraphics[width=0.9\columnwidth]{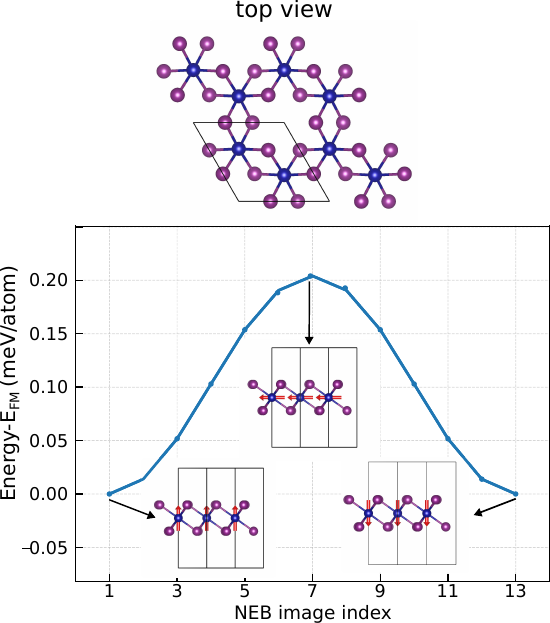}
\caption{
Collective spin reversal in monolayer CrI$_3$. Top: crystal structure viewed along
the out-of-plane direction. Bottom: minimum-energy path obtained from cDFT-NEB
calculations using 13 images for the transition between ferromagnetic states with
the Cr magnetization oriented along $+z$ and $-z$. At the midpoint of the pathway,
the magnetic moments are oriented in plane. Representative magnetic configurations
are shown for selected images.
}
\label{fig:CrI3}
\end{figure}

The calculated minimum-energy path is shown in Fig.~\ref{fig:CrI3}. The energy
profile exhibits the expected symmetric behavior, with a maximum at the midpoint
of the pathway. At this midpoint, corresponding to the seventh image, the Cr
moments are oriented along the in-plane $+y$ direction, while the induced I moments
remain antiparallel and point along $-y$. The magnitudes of the local moments remain
nearly constant during the reversal, with values of approximately $2.682~\mu_{\rm B}$
on Cr and $0.0305~\mu_{\rm B}$ on I. The resulting magnetic anisotropy barrier is
approximately $0.2$ meV/atom, representing the energy required for coherent
rotation of the magnetization from the out-of-plane easy axis toward the in-plane
direction.

This example demonstrates that the cDFT-NEB formalism is capable of describing not
only strongly coupled magnetostructural transformations, such as those encountered
in FeRh and Fe, but also transitions in which magnetic degrees of freedom dominate
the energetics. The method therefore provides a unified first-principles framework
for investigating both magnetostructural and magnetic transition pathways in a
wide range of materials.

	%%%%%%%%%%%%%%%%%%%%%%%%%%
	
	\section{Conclusion}
	%\section{4. Conclusion}
	\label{Sec:conclusion}

We have developed a generalized first-principles transition-path framework that combines  potential-based constrained  
density functional theory with a magnetostructural extension of the nudged elastic band method.
The resulting cDFT-NEB formalism enables the simultaneous optimization of atomic coordinates,
lattice degrees of freedom, and noncollinear magnetic moments within a unified variational framework.
By consistently incorporating atomic forces, stresses, and spin torques,
the method provides access to minimum-energy pathways and transition 
states in systems with coupled structural and magnetic degrees of freedom.

The implementation has been integrated into the ABINIT package and supports 
fixed- and variable-cell calculations for both collinear
and noncollinear magnetic configurations while maintaining full electronic self-consistency 
throughout the pathway optimization.

Applications to FeRh, elemental iron, and monolayer CrI$_3$ demonstrate the versatility of the method.
In FeRh, fixed- and variable-cell calculations yield similar activation barriers,
indicating a modest effect of cell relaxation on the transition pathway.
In iron, the framework captures the simultaneous lattice deformation, atomic
shuffle, volume reduction, and collapse of the magnetic moments during the
bcc-FM to hcp-NM transformation, as well as the pressure-induced reduction of
the activation enthalpy. In CrI$_3$, it enables the direct determination of
the noncollinear spin-reversal pathway and the associated magnetic anisotropy
barrier in the presence of spin-orbit coupling.

The present framework provides a general first-principles platform for investigating 
activated magnetostructural processes. It can be applied to
pressure-induced magnetic transitions, magnetovolume transformations,
noncollinear magnetic materials, and other systems governed by coupled
structural and magnetic energy landscapes.

	% Acknowledgements
	\medskip
	\textbf{Acknowledgments} \par
	
	This work is part of the SHAPEme project (EOS ID 400077525) that has received 
    funding from the FWO and F.R.S.-FNRS under the Excellence of Science (EOS)
    programme funded by the Federal government of Belgium.
    This work was supported by the Fonds de la Recherche Scientifique - F.R.S.-FNRS
    under Grant number J.0091.26 (CDR - ABINIT on GPUs : Equal footing approach).
    We thank Denis Sabani for suggesting the extension of the NEB method to include magnetic degrees of freedom.

	%\newpage
	\bibliography{main} 
	
\end{document}